\documentclass[10pt,twoside,BCOR7mm,DIV15,headinclude,footexclude,
               cleardoubleempty,idxtotoc]
{scrartcl}

\usepackage{natbib}
\usepackage[font=small,labelfont=bf]{caption}
\usepackage[english]{babel}
\usepackage{graphicx}
\usepackage{hyperref}
\usepackage{scrpage2}
\usepackage{ifthen}
\usepackage{booktabs}
\usepackage{amsmath}
\usepackage{amssymb}
\usepackage{multicol}
\usepackage{float}
\usepackage{hyperref}

\hypersetup{breaklinks=true
,colorlinks=true,linkcolor=black,urlcolor=blue
,citecolor=black}

\addto\captionsenglish{%
}

\makeatletter
\renewcommand{\@biblabel}[1]{}
\renewcommand{\@cite}[2]{%
{#1\ifthenelse{\boolean{@tempswa}}{,#2}{}}}
\makeatother
\ofoot{\thepage}
\ifoot{}

\setcapindent{0em}
\setheadsepline{1pt}

\setkomafont{pagehead}{\normalfont\sffamily}
\setkomafont{pagenumber}{\normalfont\rmfamily}

\makeatletter
\newcommand{\listofcontributions}{\@starttoc{con}}

\newcommand{\l@contribution} {\@dottedtocline{1}{1.5em}{2.3em}}
\makeatother

\newenvironment{contribution}{
\setcounter{section}{0}
\setcounter{figure}{0}
\setcounter{table}{0}
}{
\newpage
\lehead{}
\rohead{}
}

\begin{document}

\setlength{\baselineskip}{2.5ex}

\begin{contribution}

\lehead{K.\ F.\ Neugent, et al.}

\rohead{The discovery and physical parameterization of a new type of Wolf-Rayet star}

\begin{center}
{\LARGE \bf The Discovery and Physical Parameterization of a New Type of Wolf-Rayet Star}\\
\medskip

{\it\bf K.\ F.\ Neugent$^1$, P.\ Massey$^1$, D.\ J.\ Hillier$^2$, N.\ I.\ Morrell$^3$}\\

{\it $^1$Lowell Observatory, U.S.A.}\\
{\it $^2$University of Pittsburgh, U.S.A.}\\
{\it $^3$Las Campanas Observatory, Chile}

\begin{abstract}
As part of our ongoing Wolf-Rayet (WR) Magellanic Cloud survey, we have discovered 13 new WRs. However, the most exciting outcome of our survey is not the number of new WRs, but their unique characteristics. Eight of our discoveries appear to belong to an entirely new class of WRs. While one might naively classify these stars as WN3+O3V binaries, such a pairing is unlikely. Preliminary {\sc cmfgen} modeling suggests physical parameters similar to early-type WNs in the Large Magellanic Cloud except with mass-loss rates three to five times lower and slightly higher temperatures. The evolution status of these stars remains an open question.
\end{abstract}
\end{center}

\begin{multicols}{2}

\section{Discovery}
We recently began a multi-year project surveying the Magellanic Clouds for Wolf-Rayet (WR) stars. Using the Swope 1-m at Las Campanas, we took images using on- and off-band interference filters and then used a image subtraction program to identify WR candidates. The candidates were then spectroscopically confirmed using MagE on Magellan. So far we have covered $\sim$60\% of the Magellanic Clouds and discovered 13 new WRs \citep{MCWRs,MCWR15}. But, it isn't the {\it number} of newly discovered WRs that makes our survey so successful, but the {\it types} of these stars. Eight of them appear to belong to an entirely new class of WRs. 

Based on the spectrum shown in Figure~\ref{KNeugent:spectra}, one might naively classify these stars as WN3+O3V binaries. The WN3 classification comes from the star's N\,{\sc v} emission ($\lambda\lambda$4603,19 and $\lambda$4945), but lack of N\,{\sc iv}. The O3V classification comes from the strong He\,{\sc ii} absorption lines but lack of He\,{\sc i}. However such a pairing of an WN3+O3V is highly unlikely. First, O3Vs are quite rare since they are the hottest and most luminous of the dwarfs \citep{Sylvia12}. In the Large Magellanic Cloud (LMC), only around a dozen are known outside of the 30 Dor region \citep{Skiff}. Second, assuming single star evolution, the lifetimes of these stars preclude such a pairing: WN3s take $\sim$3-5 million years to form whereas a massive star evolves out of the O3V stage after only a million years. Third, we have yet to see any radial velocity variations as you would expect in a binary system, though we are still gathering data for confirmation. Fourth, our UV data (discussed later) shows no C\,{\sc iv} $\lambda$1550, the most prominent UV line in O-stars. However, the most convincing evidence deals with the stars' absolute magnitudes. The WN3/O3Vs we have observed are faint with M$_V \sim -2.5$. O3Vs, on the other hand, have much brighter absolute magnitudes with M$_V \sim -5.5$ \citep{Conti88}. Thus, we conclude these stars are not WN3+O3V binaries and adopt the naming convention of WN3/O3V.

\begin{figure}[H]
\begin{center}
\includegraphics[width=\columnwidth]{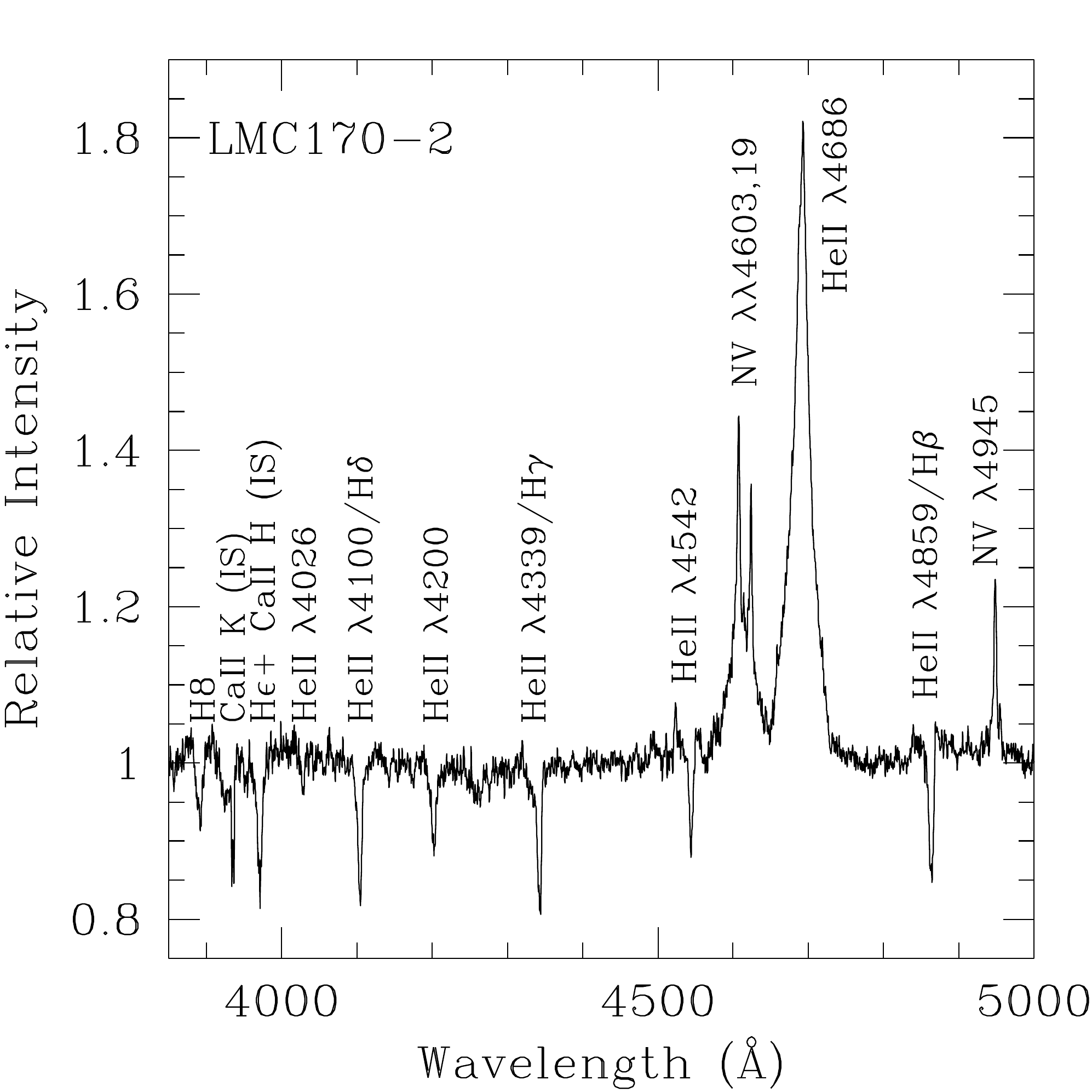}
\caption{Spectrum of LMC170-2, one of our newly discovered WN3/O3V stars \citep{MCWR15}.
\label{KNeugent:spectra}}
\end{center}
\end{figure}

In addition to the MagE spectra (resolving power, $R$, of 4100) obtained for all eight newly discovered stars, we observed three stars in the UV using COS on {\it HST}. We also observed one star using both FIRE and MIKE on Magellan, giving us the NIR and an optical spectrum with $R = 11000 - 14000$. Next we attempted to model our spectra using a single set of physical parameters.

\section{Physical Parameterization}
To model the WN3/O3s, we turned to {\sc cmfgen} \citep{CMFGEN}, a stellar atmosphere code designed for hot stars with stellar winds where the usual assumptions of plane-parallel geometry and LTE no longer hold. Using only the MagE optical data, we produced a good fit to both the emission and absorption lines in the spectra, as shown in Figure~\ref{KNeugent:modelfit}. Table~\ref{KNeugent:params} lists the physical parameters used.

\begin{figure}[H]
\begin{center}
\includegraphics[width=\columnwidth]{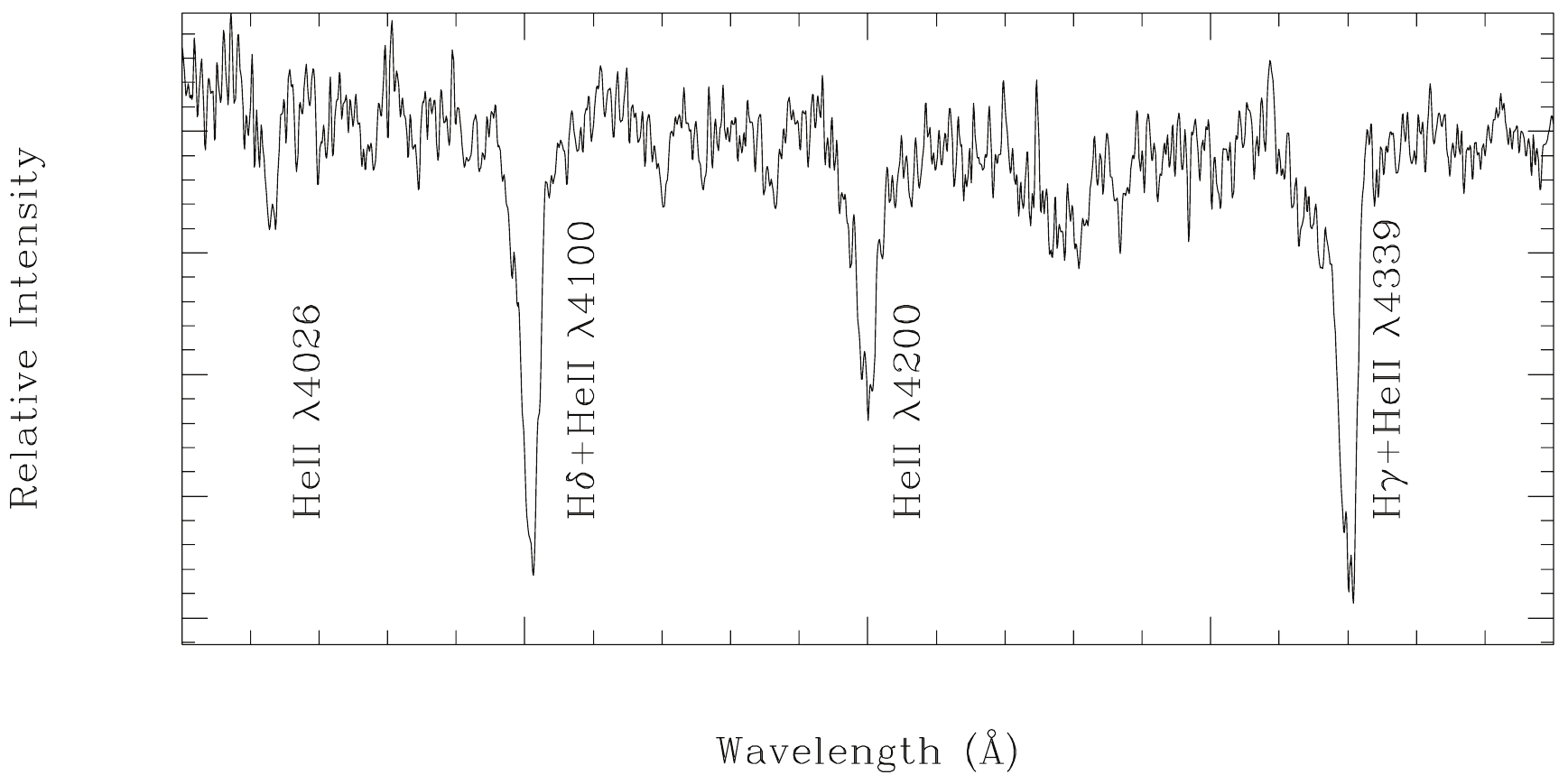}
\includegraphics[width={0.49\columnwidth}]{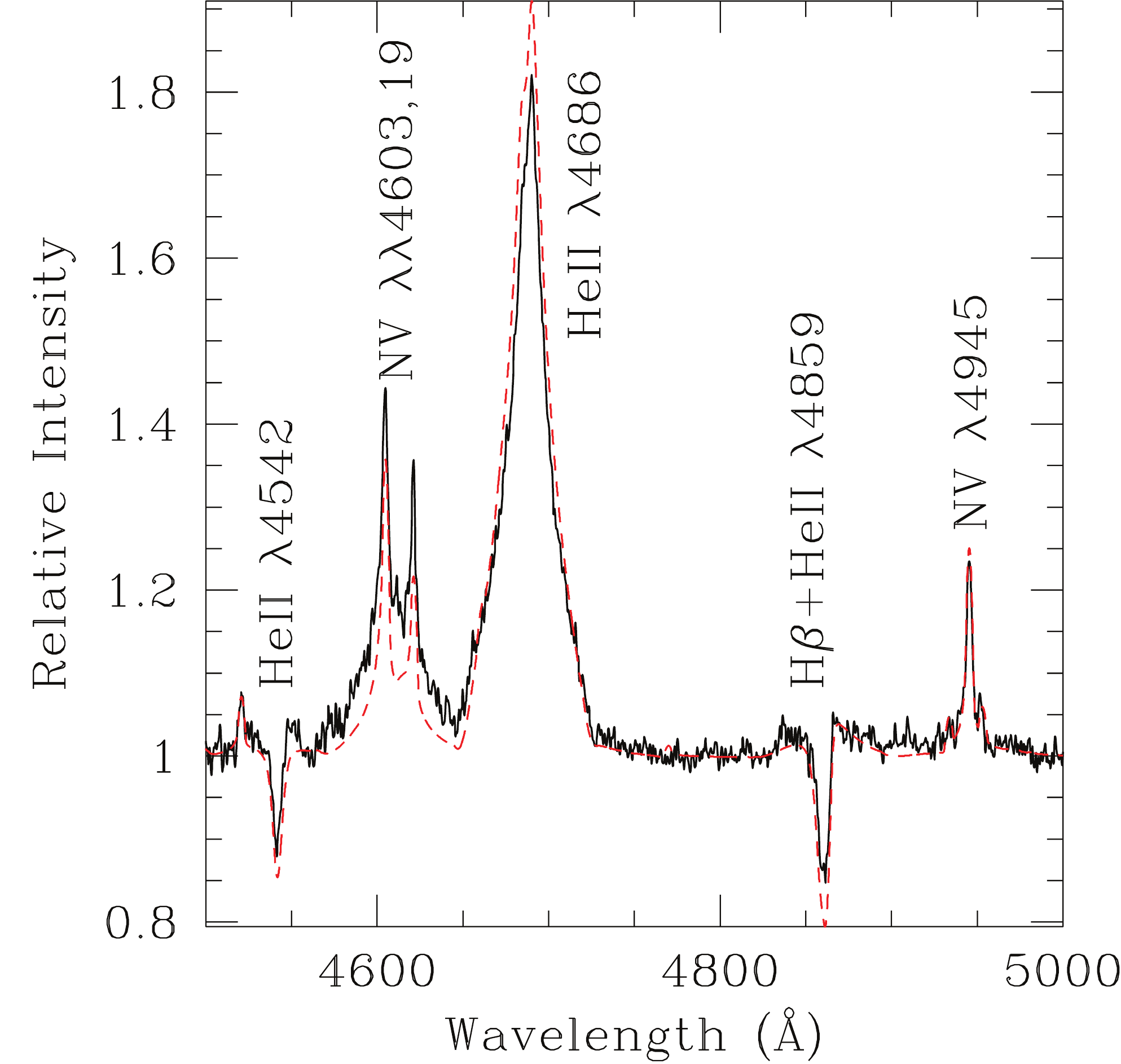}
\includegraphics[width={0.49\columnwidth}]{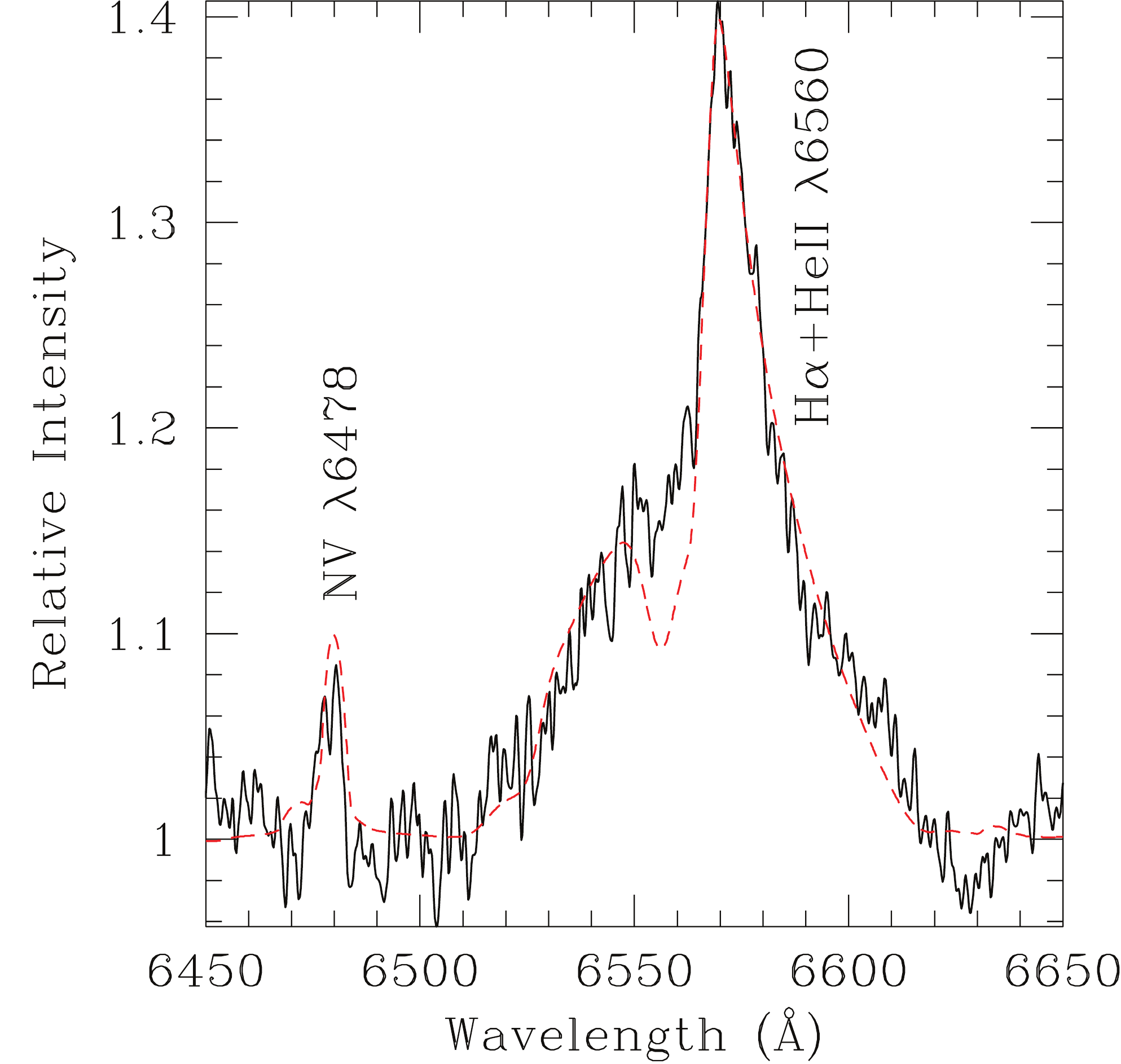}
\caption{Best CMFGEN fit to the optical spectrum of LMC170-2 from \citet{MCWRs}.
\label{KNeugent:modelfit}}
\end{center}
\end{figure}

\begin{table}[H] 
\begin{center} 
\captionabove{CMFGEN model parameters for LMC170-2} 
\label{KNeugent:params}
\begin{tabular}{lc}
\toprule
Parameter & Best Fit\\
\midrule 
$T_{\rm{eff}}$ (K) & 100,000\\
$L/L_{\odot}$ & $4\times10^{5}$\\
$\dot{M}^*$ ($M_{\odot}$/yr) & $1.2\times10^{-6}$\\
He/H (by \#) & 1.0\\
N & 10.0$\times$ solar\\
C, O & 0.05$\times$ solar\\
\bottomrule
\end{tabular}
\end{center}
\vspace{1mm}
\scriptsize{
{\it Note:} Assumes a clumping filling factor of 0.1, $v_\infty$=2400 km s$^{-1}$, $\beta$=0.8, and $v_{\rm sin i}$ = 150 km s$^{-1}$.}
\end{table}

The majority of these parameter values are comparable to those found for early-type LMC WNs \citep{Potsdam}. Despite the stars' faint visual magnitudes, their bolometric luminosities are normal. These stars appear to be evolved, with significantly enriched N and He. While their effective temperatures are a bit on the high side, they are still within the range found for other WN3s. The most unusual value is their mass-loss rate. As Figure~\ref{KNeugent:mdot} shows, our WN3/O3 star LMC170-2 has a mass-loss rate that is three to five times lower than those found for other early-type LMC WNs as analyzed by \cite{Potsdam}. As discussed later, this low mass-loss rate appears to hold true for all eight WN3/O3s currently known.

\begin{figure}[H]
\begin{center}
\includegraphics[width={0.85\columnwidth}]{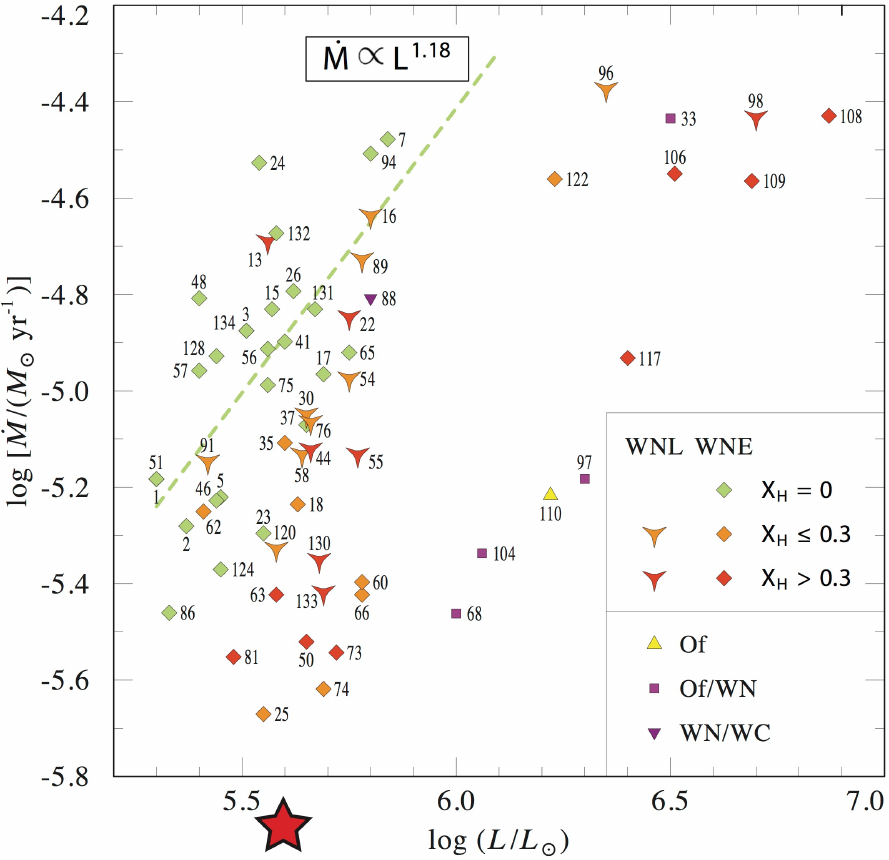}
\caption{Luminosity vs.\ mass-loss rate for LMC170-2 (large red star) compared to those of other early-type LMC WNs analyzed by \citet{Potsdam}. Adapted from Fig.\ 6 of \citet{Potsdam} and used with permission.
\label{KNeugent:mdot}}
\end{center}
\end{figure}

Since the physical parameters for this model were determined using only the optical data, we were curious to see how well the model would match our newly obtained UV and NIR spectra. As is shown in Figure~\ref{KNeugent:UVNIR}, the agreement between the model and our UV and NIR spectra is practically perfect. Due to space constraints we have only shown the model fit for He\,{\sc ii}$\lambda$1640 in the UV, but it should be noted that all of our diagnostic lines such as the O\,{\sc iv}$\lambda$1038 resonance doublet, N\,{\sc v}$\lambda$1240, and the lack of C\,{\sc iv}$\lambda$1550 are similarly well fit. As an added bonus, the spectral energy distribution of the model almost perfectly matches our UV, optical and NIR spectra. This confirms that the reddening is well determined. However, it should be noted that for stars with $T_{\rm eff} >$ 30,000~K, the spectral energy distributions will all look similar longwards of 1000\AA\ as this is the Rayleigh-Jeans tail of the black body distribution.

\begin{figure}[H]
\begin{center}
\includegraphics[width={0.49\columnwidth}]{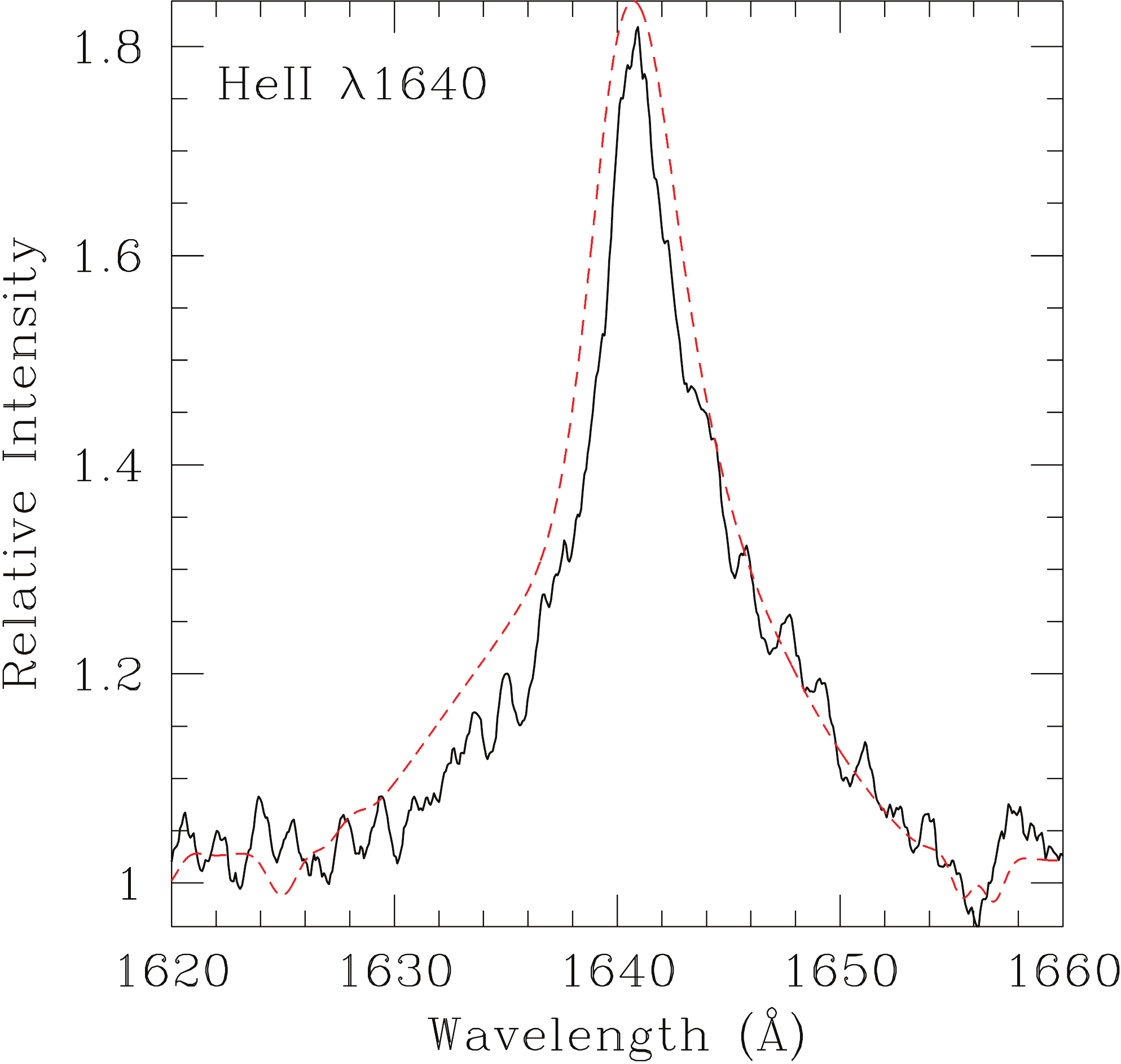}
\includegraphics[width={0.49\columnwidth}]{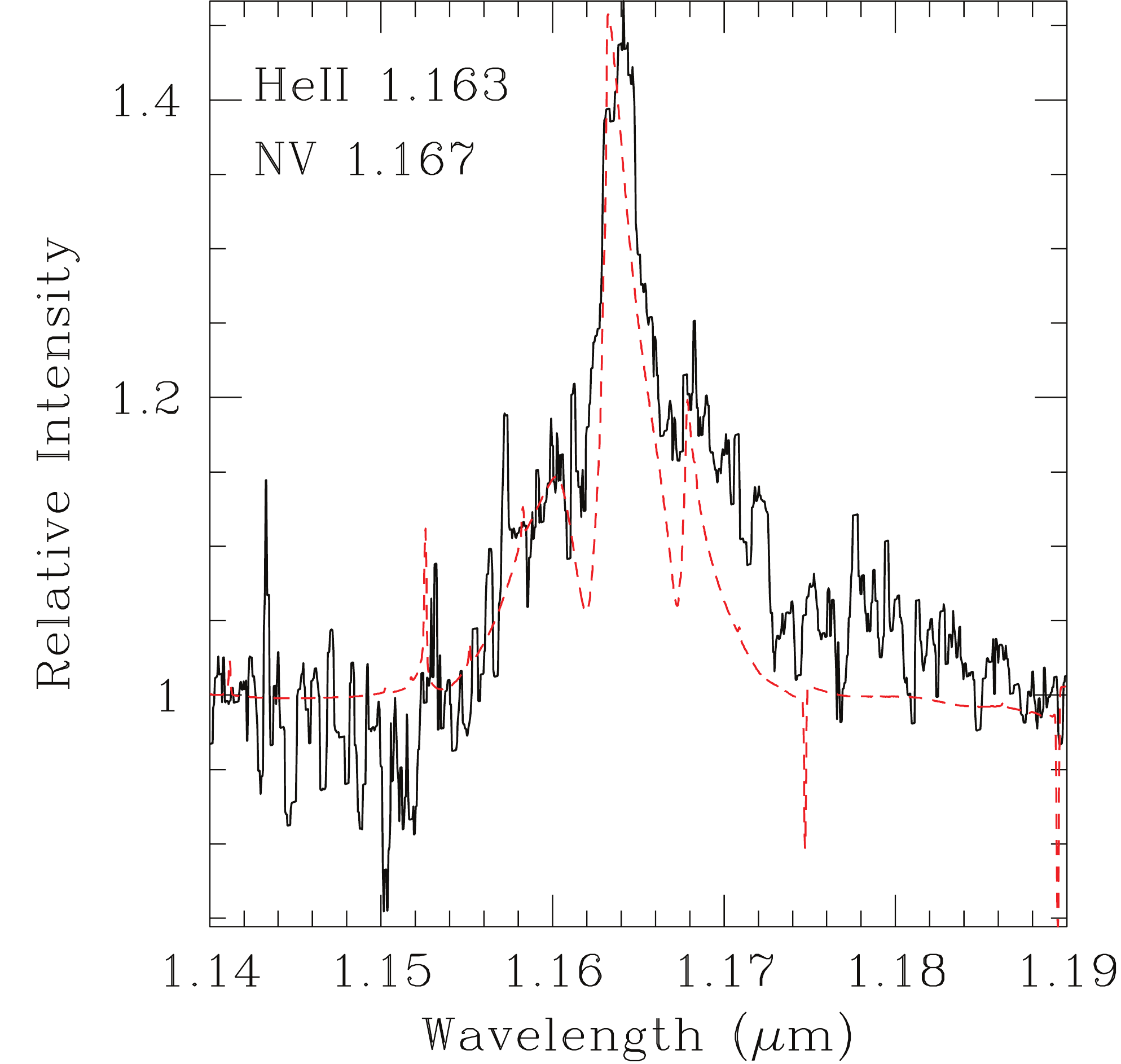}
\caption{Best CMFGEN fit to the UV and NIR spectrum of LMC170-2.
\label{KNeugent:UVNIR}}
\end{center}
\end{figure}

With our exemplary model in hand, we began a sensitivity analysis of the model's input parameters. We first attempted to slightly raise and lower the value of each parameter (e.g., increase by 10\%, then decrease by 10\%). As co-author John Hillier wrote, we soon discovered that, ``it's an interesting regime -- $\dot{M}$, H/He, $T_{\rm eff}$, and log g all seem to interact in complex ways." For example, decreasing $T_{\rm eff}$ from 100,000 K to 90,000 K yielded a clearly inferior fit to the He and H lines. However, increasing $T_{\rm eff}$ to 110,000 K brought the model too close to the Eddington limit. This could be avoided by changing the surface gravity in parallel with $T_{\rm eff}$. So, doing a clear-cut sensitivity analysis proved to be more difficult than originally expected. Still, we were able to get reasonable uncertainty estimates for the majority of the parameters. However, we hope to refine these once we have separately modeled all eight of the spectra. 

Figure~\ref{KNeugent:WN3O3} shows the striking similarities between the spectra of our WN3/O3s, suggesting that a good model fit to one star will also fit the remaining seven stars quite nicely. And indeed, this is what we found. While there are slight differences, we expect that the physical parameters will only change by small amounts for each star. Certainly none of the spectra suggest a drastically different temperature or mass-loss rate (for example).
\begin{figure}[H]
\begin{center}
\includegraphics[width=\columnwidth]{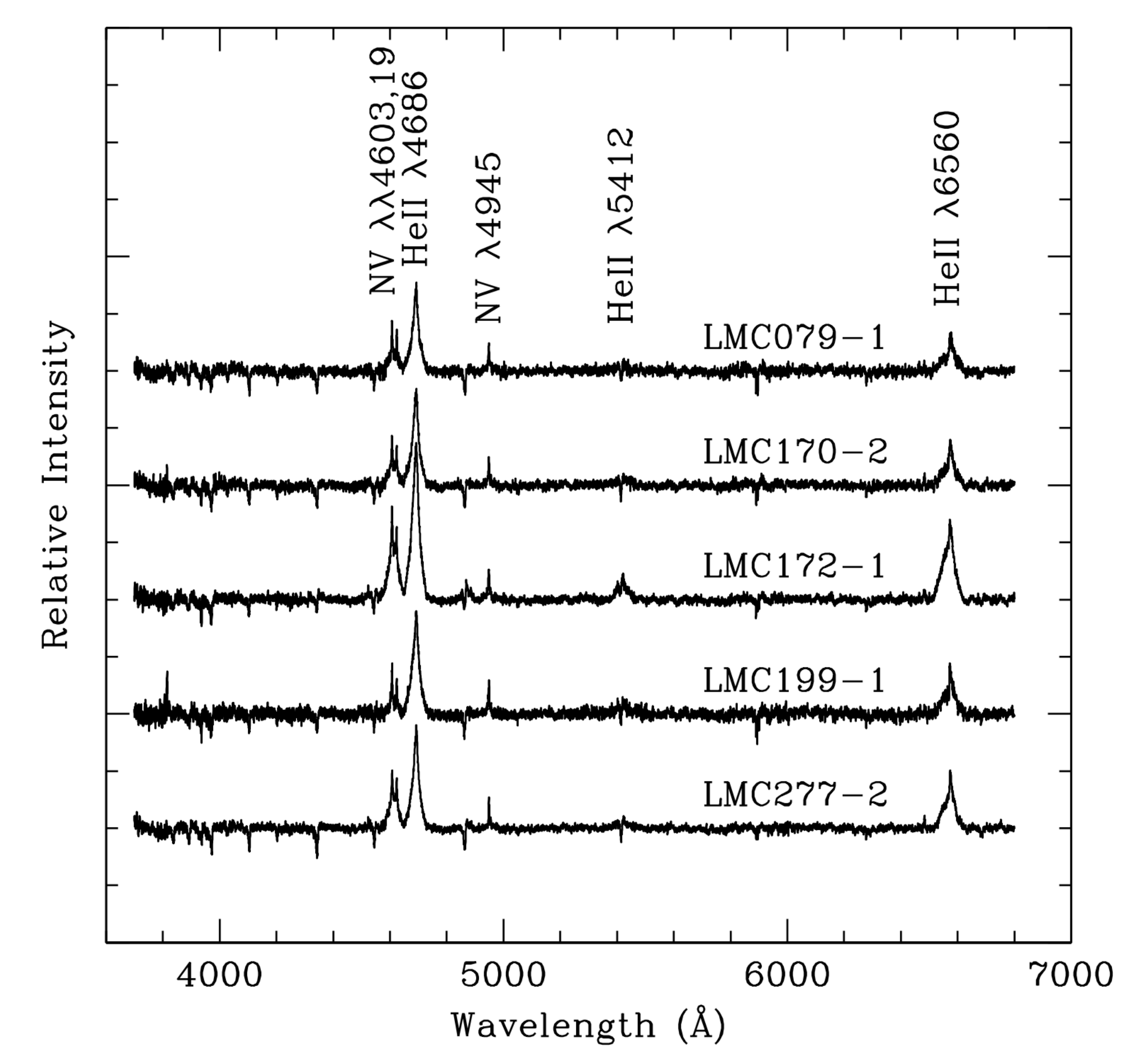}
\caption{Spectra of five of our newly discovered WN3/O3s from \citet{MCWRs}.
\label{KNeugent:WN3O3}}
\end{center}
\end{figure}

\section{Next Up}
We still have a lot to learn about these objects and certainly there are many questions left to be answered. We are currently working on constraining the model parameters and determining whether the ``best fit" model presented above is unique or if a similarly good model can be obtained using a different set of parameters. Once this has been determined, we will continue modeling all eight stars to get a more global view of the range of expected parameter values within this WR subtype. At that point we can begin working with our stellar evolutionary colleagues to determine whether current models predict the existence of these stars. 

We additionally hope to get a better handle on where these stars come from. If they were spatially separated, or more isolated, from the ``normal" LMC WRs, one might infer that the WN3/O3s were formed from different (lower metallicity?) progenitors. However their spatial distribution suggests that this is not the case. We also haven't ruled out an explanation dealing with binary evolution. While they don't appear to be binaries now, it is possible that they were in the past. As we continue on with our WR survey in the Magellanic Clouds, we hope to find even more of these exciting new stars. Within the next year we should have a much better grasp of their numbers, physical parameters and possibly even their origins. 

This work has been supported by the National Science Foundation under AST-1008020. Support for program number HST-GO-13780 was provided by NASA through a grant from the Space Telescope Science Institute, which is operated by the Association of Universities for Research in Astronomy, Incorporated, under NASA contract NAS5-2655. K.F.N. received a National Science Foundation travel grant from the American Astronomical Society that funded her airline travel to and from the meeting in Potsdam. D.J.H. also acknowledges support from STScI theory grant HST-AR-12640.01.

\bibliographystyle{aa} 
\bibliography{myarticle}

\begin{thebibliography}{7}
\expandafter\ifx\csname natexlab\endcsname\relax\def\natexlab#1{#1}\fi

\bibitem[{{Conti}(1988)}]{Conti88}
{Conti}, P.~S. 1988, NASA Special Publication, 497, 119

\bibitem[{{Ekstr{\"o}m} {et~al.}(2012){Ekstr{\"o}m}, {Georgy}, {Eggenberger},
  {Meynet}, {Mowlavi}, {Wyttenbach}, {Granada}, {Decressin}, {Hirschi},
  {Frischknecht}, {Charbonnel}, \& {Maeder}}]{Sylvia12}
{Ekstr{\"o}m}, S., {Georgy}, C., {Eggenberger}, P., {et~al.} 2012, \aap, 537,
  A146

\bibitem[{{Hainich} {et~al.}(2014){Hainich}, {R{\"u}hling}, {Todt}, {Oskinova},
  {Liermann}, {Gr{\"a}fener}, {Foellmi}, {Schnurr}, \& {Hamann}}]{Potsdam}
{Hainich}, R., {R{\"u}hling}, U., {Todt}, H., {et~al.} 2014, \aap, 565, A27

\bibitem[{{Hillier} \& {Miller}(1998)}]{CMFGEN}
{Hillier}, D.~J. \& {Miller}, D.~L. 1998, \apj, 496, 407

\bibitem[{{Massey} {et~al.}(2015){Massey}, {Neugent}, \& {Morrell}}]{MCWR15}
{Massey}, P., {Neugent}, K.~F., \& {Morrell}, N. 2015, \apj, 807, 81

\bibitem[{{Massey} {et~al.}(2014){Massey}, {Neugent}, {Morrell}, \&
  {Hillier}}]{MCWRs}
{Massey}, P., {Neugent}, K.~F., {Morrell}, N., \& {Hillier}, D.~J. 2014, \apj,
  788, 83

\bibitem[{{Skiff}(2014)}]{Skiff}
{Skiff}, B.~A. 2014, VizieR Online Data Catalog, 1, 2023

\end{thebibliography}

\end{multicols}

\end{contribution}


\end{document}